\documentclass[conference]{IEEEtran}
\usepackage[T1]{fontenc}
\usepackage{graphicx} 
\usepackage{footnote}
\usepackage{url}
\usepackage[table]{xcolor}
\usepackage{xcolor}
\usepackage{listings}
\usepackage{tabularx}
\usepackage{multirow}
\usepackage{booktabs}
\usepackage{multirow}
\usepackage{algorithm}
\usepackage{algpseudocode}
\usepackage[flushleft]{threeparttable}
\usepackage[colorinlistoftodos]{todonotes}
\usepackage{enumitem}
\usepackage{caption}
\usepackage{xurl}

\definecolor{delim}{RGB}{20,105,176}
\definecolor{numb}{RGB}{106, 109, 32}
\definecolor{string}{RGB}{0, 0, 0}
\definecolor{punct}{RGB}{128, 128, 128}
\lstdefinelanguage{json}{
    showspaces=false,
    showtabs=false,
    breaklines=true,
    postbreak=\raisebox{0ex}[0ex][0ex]{\ensuremath{\color{gray}\hookrightarrow\space}},
    breakatwhitespace=true,
    basicstyle=\ttfamily\small,
    upquote=true,
    morestring=[b]",
    stringstyle=\color{string},
    literate=
     {0}{{{\color{numb}0}}}{1}
      {1}{{{\color{numb}1}}}{1}
      {2}{{{\color{numb}2}}}{1}
      {3}{{{\color{numb}3}}}{1}
      {4}{{{\color{numb}4}}}{1}
      {5}{{{\color{numb}5}}}{1}
      {6}{{{\color{numb}6}}}{1}
      {7}{{{\color{numb}7}}}{1}
      {8}{{{\color{numb}8}}}{1}
      {9}{{{\color{numb}9}}}{1}
      {:}{{{\color{punct}{:}}}}{1}
      {,}{{{\color{punct}{,}}}}{1}
      {\{}{{{\color{delim}{\{}}}}{1}
      {\}}{{{\color{delim}{\}}}}}{1}
      {[}{{{\color{delim}{[}}}}{1}
      {]}{{{\color{delim}{]}}}}{1},
}

\lstdefinelanguage{JavaScript}{
  keywords={typeof, new, true, false, catch, function, return, null, catch, switch, var, if, in, while, do, else, case, break, const},
  keywordstyle=\color{blue}\bfseries,
  ndkeywords={class, export, boolean, throw, implements, import, this, it, expect, test, require},
  ndkeywordstyle=\color{darkgray}\bfseries,
  identifierstyle=\color{black},
  sensitive=false,
  comment=[l]{//},
  morecomment=[s]{/*}{*/},
  commentstyle=\color{purple}\ttfamily,
  stringstyle=\color{red}\ttfamily,
  morestring=[b]',
  morestring=[b]"
}

\newcommand{\Fix}[1]{\textcolor{black}{#1}}
\newcommand{\add}[1]{\textcolor{black}{#1}}
\newcommand{\Space}[1]{}

\newcommand{\NumInitialProjects}{\Fix{245}}
\newcommand{\NumProjectsTestNode}{\Fix{170}}
\newcommand{\NumProjects}{\Fix{100}}
\newcommand{\NumProjectsWithFailure}{\Fix{66}}
\newcommand{\NumProjectsBrowser}{\Fix{16}}
\newcommand{\NumEnvFlakyTests}{1355}

\newcommand{\NumProjectsFinalDataset}{\Fix{116}}

\newcommand{\NumConflicts}{\Fix{12}}
\newcommand{\NumNoConflicts}{\Fix{56}}
\newcommand{\NumRerun}{10}

\newcommand{\EnvFlakiness}{environmental flakiness}
\newcommand{\EnvFlakyProjects}{environmental flaky projects}
\newcommand{\EnvDepProjects}{environmental dependent projects}
\newcommand{\EnvDepTests}{environmental dependent tests}

\newcommand{\EnvFlakyTests}{environmental flaky tests}
\newcommand{\EnvFlakyTest}{environmental flaky test}
\newcommand{\FlakyProjects}{flaky projects}

\newcommand{\NumPassed}{50}
\newcommand{\NumEnvFlaProjects}{39}
\newcommand{\NumEnvFlaTests}{28}
\newcommand{\NumEnvProjects}{65}
\newcommand{\NumFlaky}{five}

\newcommand{\NumEnvFlaNode}{five}
\newcommand{\NumEnvFlaOS}{28}
\newcommand{\NumEnvFlaBrowser}{17}
\newcommand{\NumEnvFlaOSNode}{16}

\newcommand{\NumEnvFlaProjectsNode}{three}
\newcommand{\NumEnvFlaProjectsOS}{21}
\newcommand{\NumEnvFlaProjectsBrowser}{eight}
\newcommand{\NumEnvFlaProjectsOSNode}{eight}

\newcommand{\NumEnvFlaTestsNode}{two}
\newcommand{\NumEnvFlaTestsOS}{seven}
\newcommand{\NumEnvFlaTestsBrowser}{nine}
\newcommand{\NumEnvFlaTestsOSNode}{eight}

\newcommand{\NEDP}{Non-flaky projects}
\newcommand{\EDP}{Environment-dependent projects}
\newcommand{\EFP}{Environmental flaky projects}
\newcommand{\EFT}{Environmental flaky tests}
\newcommand{\FP}{Flaky projects}

\newcommand{\tool}{\texttt{js\textnormal{-}env\textnormal{-}sanitizer}}

\newcommand{\NumEvaluationProjects}{\Fix{15}}
\newcommand{\NumMixFrameworks}{\Fix{six}}
\newcommand{\NumEvaluationProjectsFinal}{\Fix{seven}}

\title{A Systematic Evaluation of Environmental Flakiness in JavaScript Tests}
\author{
    \IEEEauthorblockN{Negar Hashemi\IEEEauthorrefmark{1}, Amjed Tahir\IEEEauthorrefmark{1}, August Shi\IEEEauthorrefmark{2}, Shawn Rasheed\IEEEauthorrefmark{3},  Rachel  Blagojevic\IEEEauthorrefmark{1}}
    \IEEEauthorblockA{\IEEEauthorrefmark{1}Massey University, New Zealand
    \\}
    \IEEEauthorblockA{\IEEEauthorrefmark{3}Victoria University of Wellington, New Zealand
  }

    \IEEEauthorblockA{\IEEEauthorrefmark{2} The University of Texas at Austin, USA
   }
}

\makeatletter
\newcommand\fs@betterruled{%
  \def\@fs@cfont{\bfseries}\let\@fs@capt\floatc@ruled
  \def\@fs@pre{\vspace*{5pt}\hrule height.8pt depth0pt \kern2pt}%
  \def\@fs@post{\kern2pt\hrule\relax}%
  \def\@fs@mid{\kern2pt\hrule\kern2pt}%
  \let\@fs@iftopcapt\iftrue}
\floatstyle{betterruled}
\restylefloat{algorithm}
\makeatother

\begin{document}

\maketitle


\begin{abstract}
Test flakiness is a significant issue in industry, affecting test efficiency and product quality.
While extensive research has examined the impact of flaky tests, many root causes remain unexplored, particularly in the context of dynamic languages such as JavaScript.
In this paper, we conduct a systematic evaluation of the impact of environmental factors on test flakiness in JavaScript.
We first executed test suites across multiple environmental configurations to determine whether changes in the environment could lead to flaky behavior. 
We selected three environmental factors to manipulate: the operating system, the Node.js version, and the browser.
We identified a total of \NumEnvProjects{} \EnvFlakyProjects{}, with \NumEnvFlaOS{} related to operating system issues, \NumEnvFlaNode{} to Node.js version compatibility, \NumEnvFlaOSNode{} to a combination of operating system and Node.js issues, and \NumEnvFlaBrowser{} related to browser compatibility. 
To address environmental flakiness, we developed a lightweight mitigation approach, \tool{}, that can \textit{sanitize} environmental-related flaky tests by skipping and reporting them (rather than failing), allowing CI builds to continue/succeed without rerunning entire test suites.
The tool achieves high accuracy with minimal performance or configuration overhead, and currently supports three popular JavaScript testing frameworks (Jest, Mocha, and Vitest).

\end{abstract}

\maketitle

\section{Introduction}
\label{sec:introduction}

Flaky tests are tests whose outcomes alternate between pass and fail without any changes to the code.
They are recognized as a phenomenon that continuously causes significant problems in industry~\cite{alshahwan2023software,Machalica2020Probabilistic} and negatively impacts product quality and delivery~\cite {fowler2011eradicating,palmer2019}, as well as testing-related activities~\cite{machalica2019predictive,lam2020dependent}.
Several reports from tech companies highlighted the high cost and impact of flaky tests on their projects, including reports from Google~\cite{ziftci2020flake}, Microsoft~\cite{Improvin2:online}, and GitHub~\cite{Reducing45:online}.   

\emph{Environmental flakiness} refers to test failures that are caused not by issues in the test logic or application code itself, but by variations in the environment in which the tests are executed.
Environmental factors may include any variations in the operating systems (OS), hardware configurations, available system resources, or browsers~\cite{amjed2022review}.
When a test passes in one environment but fails in another without any code changes, it is often considered to be environmentally flaky.
This type of flakiness is particularly challenging to detect, as it may depend on subtle, non-deterministic interactions with the system.
For example, a test might fail intermittently on macOS but pass consistently on Linux, or it might behave differently when executed in a container than in a native environment. 

As modern software systems increasingly rely on distributed, containerized, and cross-platform deployments, managing environmental variability becomes critical to ensuring stable and trustworthy test results. Test consistency across environments is important as any flakiness resulting from environmental variation can undermine the reliability of continuous integration (CI) pipelines, delay deployments, and introduce unnecessary engineering effort~\cite{thorve2018empirical}.

Previous studies have identified and categorized potential causes of flaky tests across different programming languages, including Java~\cite{luo2014empirical}, Python~\cite{gruber2021empirical}, and JavaScript~\cite{Hashemi2022flakyJS}.
Hashemi et al.~\cite{Hashemi2022flakyJS} reported the first extensive empirical study on flaky tests in open-source JavaScript projects, categorizing the causes and fixing strategies for these flaky tests.
They found that one of the major causes of flakiness in JavaScript is the environment in which the tests are run (a similar observation was reported in Costa et al.~\cite{costa2022test}).  

In this study, we conducted an in-depth investigation into \EnvFlakiness{} in JavaScript, focusing on the OS, browsers, and Node.js versions.
Our goal is to understand the prevalence of flakiness by changing the environmental factors.
We also want to understand the reasons for the failure after running tests in various environment configurations.
Based on our empirical results, we aim to develop a technique that can help developers mitigate \EnvFlakiness{}.
In practice, developers are reported to employ several methods to address test flakiness, such as adding suitable assumptions to the tests to run them in specific environments or with specific configurations~\cite{Hashemi2022flakyJS,amjed2022review}.
Developing an automated technique to sanitize tests against environmental conditions can help developers gain more control over the tests' behavior.

The main contributions of this paper are the answers to the following research questions:\\
 \noindent  \textbf{RQ1} How prevalent are \EnvFlakyTests{} in JavaScript?\\
 \noindent  \textbf{RQ2} What are the root causes of \EnvFlakiness{}?\\
 \noindent  \textbf{RQ3} How can we mitigate test flakiness caused by environmental factors? 


RQ1 aims to investigate the prevalence of \EnvFlakiness{} in JavaScript.
We employed a systematic rerun approach to uncover \EnvFlakyTests{}.
By changing environmental conditions and rerunning the tests under different conditions, we assess whether the test outcomes change.
RQ2 aims to understand the main causes of \EnvFlakyTests{} in JavaScript.
We manually inspected each detected \EnvFlakyTest{} and classified the cause of flakiness into different categories.
RQ3 aims to identify a mitigation method for flaky tests caused by environmental variation.

Based on our systematic evaluation of possible \EnvFlakiness{} in the \NumProjectsFinalDataset{} projects we investigated, we identified a total of \NumEnvFlaProjects{} \EnvFlakyProjects{} (i.e., flakiness that occurs at the project level due to specific project setups or package dependencies).
We also identified \NumEnvFlaTests{} \EnvFlakyTests{} (i.e., tests that fail due to the change in the test environment (OS, Node.js version and/or the browser)).
Looking more closely at the causes, we found that \NumEnvFlaOS{} flaky tests are related to operating system issues, \NumEnvFlaNode{} are related to Node.js version dependencies, and \NumEnvFlaBrowser{} are related to browser setup.
To mitigate such \EnvFlakiness{} causes by variation of OS, Node.js, or browsers, we developed an annotation-based approach, implemented in a tool called \tool{}, that can sanitize and quarantine such flaky tests by skipping and reporting them (instead of failing the tests), which allows CI builds to continue/succeed without the need to rerun entire test suites or stopping the build for developers to inspect (flaky) failures.
Our tool is framework-independent and works as a plugin that currently supports three popular JavaScript testing frameworks (Jest~\cite{Jest}, Mocha~\cite{Mochathe31}, and Vitest~\cite{VitestNe17}).
We uncovered that \tool{} helps skip and report all tests that typically fail due to environmental variation.

\section{Background}
\label{sec:background}
The environment in which programs are executed can significantly influence their behavior, which, in turn, can affect test execution and lead to changes in test outcomes depending on the environment configuration.
These tests are therefore \emph{flaky} because their outcomes depend not only on the code they test but on external environmental factors. \add{We refer to such tests as \emph{environmental flaky tests}: tests that fail in one environment but pass in others (with other variables in control, including code). These failures may be consistently tied to a specific environment or occur intermittently. We consider compatibility issues as a subset of environmental flakiness, as they manifest as failures that occur only in certain environments (e.g., due to differences in API implementations across JVMs, while the calling code remains unchanged~\cite{shi2016nondex}).}

\subsection{Environmental Factors}
In the context of our study, we considered three environmental factors that may contribute to test flakiness in JavaScript: (1)~the OS, (2)~the version of the runtime environment (Node.js), and (3)~the browsers.

\subsubsection{Operating Systems}
OSs are a major cause of flakiness in JavaScript~\cite{Hashemi2022flakyJS}, with Ubuntu, macOS, and Windows being the most commonly used platforms~\cite{adekotujo2020comparative}.
Differences between OS can lead to inconsistent test behavior due to variations in properties such as file systems (e.g., path separators, case sensitivity), line endings, default locale and encoding, performance and timing, and permission models~\cite{amjed2022review, job2025and}. 


\subsubsection{Node.js}
Node.js~\footnote{\url{https://nodejs.org/en}} is a cross-platform JavaScript runtime environment popular in the JavaScript community~\cite{stackoverflow2024survey}.  
Node.js allows developers to execute JavaScript code outside of a web browser.
It is widely adopted in open-source and enterprise projects, with a rich ecosystem of packages supported through the Node Package Manager (npm). 
Node.js follows semantic versioning (SemVer), where major releases can introduce breaking changes. With the numerous available Node.js versions, it is important to consider how these different versions may impact test outcomes.  
The JavaScript engine and built-in APIs may change and update between Node.js versions.
These changes can affect how code executes, especially in terms of timing and asynchronous behavior, which may expose hidden race conditions or timing-related issues in tests.

\subsubsection{Browsers}
Browsers introduce another source of environmental variability for tests.
A large number of JavaScript projects involve processing and rendering browser-related events.
Different browsers make slightly different choices about rendering, timing, caching, and permissions.
These differences can cause a test to pass in one browser but intermittently fail in another.
In addition, the interaction between OS and browser can further contribute to test flakiness, as they may behave differently depending on the underlying OS.  
For example, a browser feature might work reliably on macOS but behave inconsistently on Windows due to differences in OS-level APIs or hardware acceleration support.

\subsection{GitHub Actions}
We utilised GitHub Actions to run our experiment, enabling us to systematically explore the impact of environmental variations on test outcomes.
GitHub Action\footnote{\url{https://docs.github.com/en/actions}} is a powerful automation tool integrated directly into GitHub that enables developers to define and execute custom workflows for CI~\cite{githubActionsCICD}.
With a GitHub Actions workflow, developers can automate tasks such as building, testing, and deploying code whenever changes are pushed to a repository.
GitHub Actions supports a variety of OS, including Ubuntu, macOS, and Windows, and it provides access to pre-installed environments with popular development tools and languages. 
For our experiments, we set up a GitHub Actions workflow to run tests from each project across multiple OS (Ubuntu, macOS, and Windows), Node.js versions, and browsers in consistent, isolated environments.

\section{Methodology}
\label{sec:methodology}
\begin{figure*}[h]
    \centering
    \includegraphics[width=0.9\linewidth]{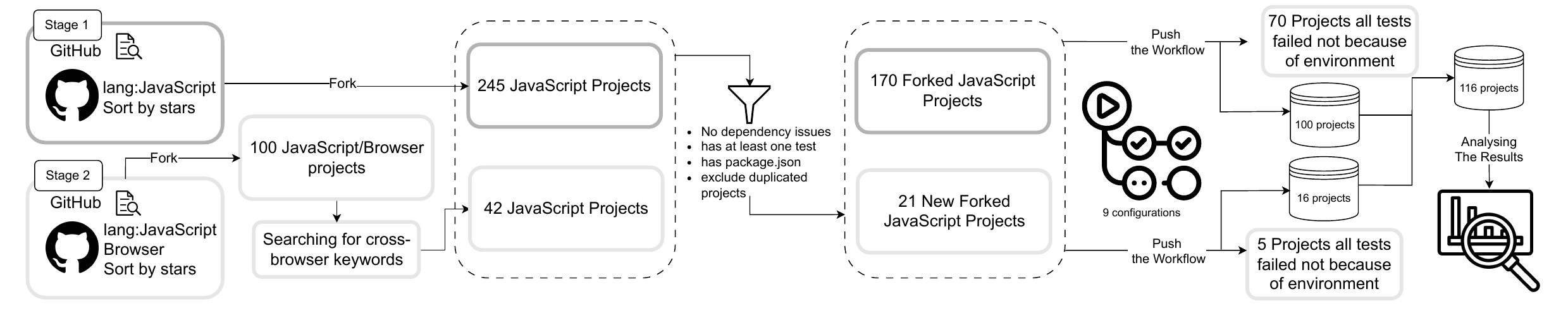}
    \caption{An overview of the data collection and analysis process}
    \label{fig:method}
\end{figure*}

We conducted our experiment in two phases: (1)~we ran tests from each project under different environment configurations to see whether tests have different outcomes, and then (2)~we manually analyzed the flaky failures and categorized the causes of flakiness.
Figure~\ref{fig:method} shows an overview of our experiment, including the data collection and analysis process.

\subsection{Dataset}
\label{sec:dataset}
We constructed a dataset of JavaScript projects by using GitHub's REST API~\cite{RESTAPIe8:online} to collect the most popular projects ranked by GitHub stars.
GitHub stars are a metric that indicates the level of interest a project has received, with prior work finding a correlation between the number of stars and the number of forks and contributors~\cite{brisson2020we}.
We selected the \NumInitialProjects{} most starred JavaScript projects to form a large enough initial dataset for inspection. 
We then filtered out projects that do not have any tests or do not utilize the Node.js runtime environment, which resulted in \NumProjectsTestNode{} JavaScript projects.
Note that these projects may use different JavaScript testing frameworks (e.g., Jest, Mocha, and Vitest).
 
After we forked the \NumProjectsTestNode{} projects, we committed and pushed a GitHub Actions workflow into our forked repository to allow us to run these tests ourselves in different environments using the GitHub Actions CI.
We noted that some runs can fail because of dependency errors or blocked running actions in the projects.
We manually customized the workflow to fix dependency errors or failed tests. \add{For example, some projects could not be installed without using the \texttt{—force} flag, while others required a specific library/tool version (e.g., Python 2) to be installed that was not available by default.}
We excluded any projects where we could not build or manually fixed, 
leaving us with \NumProjects{} projects in our dataset.

Since one of the primary applications of JavaScript is for scripting in web browsers, we aim to investigate how the combination of OS and browsers contributes to test flakiness. 
However, we find that our dataset contained only 22 web-based JavaScript projects. Therefore, 
we conducted an additional search on GitHub to identify and locate projects specifically designed to run cross-browser tests.
By filtering for projects containing the keyword ``browser'', we selected the 100 most-starred repositories to expand our dataset.
To determine whether a project includes cross-browser testing, we examined each project's package file for browser-specific keywords.
We selected a set of terms commonly associated with tools and frameworks used for browser-based or end-to-end testing, including: ``playwright'', ``karma-chrome-launcher'', ``karma-firefox-launcher'', ``karma-safari-launcher'', ``selenium-webdriver'', ``webdriverio'', ``testcafe'', ``cypress'', ``test:e2e'', ``gulp'' and``grunt''.
This search identified 42 projects, 11 of which were already available in our initial dataset. We apply the same processing steps used for the initial dataset to the remaining 31 projects.
After excluding overlapping projects between the two datasets, we construct an additional dataset consisting of \NumProjectsBrowser{} new projects.
Overall, our combined dataset contains \NumProjectsFinalDataset{} projects.
Table~\ref{tab:dataset} summarizes the statistics of the projects in our dataset.

\begin{table}[!htp]\centering
\caption{Project Statistics}\label{tab:dataset}

\begin{tabular}{l|rrrr}\toprule
Stats & \#Stars    & \#Forks   & \#Contributors & \#Test Files \\\midrule
Mean                  & 35411.27 & 5482.19 & 314.78       & 63.72      \\
Total                 & 4107707  & 635934  & 36515        & 7392   
      \\
\bottomrule
\end{tabular}

\end{table}

\subsection{Procedure}
With our dataset of JavaScript projects, we set up a procedure that can run tests from those projects to check whether variance in environmental factors can change test outcomes.
For each project, we created a GitHub Actions workflow file to run tests in different environments.
In this file, we define different combinations of environmental factors (Listing~\ref{lst:yml}).
This file enables the GitHub CI pipeline to run each project across nine distinct environment configurations, with each configuration rerun \NumRerun{} times, resulting in a total of 90 runs. 

\begin{lstlisting}[caption={Different environmental setups in GitHub Actions},label={lst:yml}]
strategy:
      fail-fast: false
      matrix:
        os: [ubuntu-latest, macos-latest, windows-latest] # Operating systems to test
        node-version: [18, 20, 22]                      # Node.js versions to test
        attempt: [1, 2, 3, 4, 5, 6, 7, 8, 9,10]         # Repeat tests 10 times

 \end{lstlisting}

Table~\ref{tab:github-actions-defaults} shows the list of the selected OS, Node.js versions, and browsers available for GitHub Actions\footnote{\url{https://github.com/actions/runner-images\#available-images}}.
For OS, we use the latest versions available on GitHub Actions.
For Node.js, we select the latest stable releases (versions 18, 20, and 22).
Ultimately, we pair the different versions of OS and Node.js to form configurations.
The three primary OSs (Ubuntu 24.04, macOS 14, and Windows Server 2022) each include multiple pre-installed browsers, such as Google Chrome, Microsoft Edge, Mozilla Firefox, and Safari (exclusive to macOS).
Browser versions vary slightly across these OS, affecting feature support and rendering behavior.
For instance, Safari is available only on macOS, whereas Microsoft Edge and Firefox are available on both Ubuntu and Windows. 
We therefore cannot explicitly change browser versions, as they are tightly coupled with the OS.


\begin{table}[]
\caption{Default OS, Node.js, and Browser in GitHub Actions runners (at the time of running the experiment)}
\label{tab:github-actions-defaults}
\resizebox{\columnwidth}{!}{%
\begin{tabular}{@{}llll@{}}
\toprule
\textbf{Label} & \textbf{OS} & \textbf{Node.js} & \textbf{Default Browser} \\ \midrule
\multirow{3}{*}{ubuntu-latest} & \multirow{3}{*}{Ubuntu 24.04} & 18.20.8 & Google Chrome 136.0.7103.92 \\
 &  & 20.19.1 & Microsoft Edge 136.0.3240.64 \\
 &  & 22.15.0 & Mozilla Firefox 138.0.1 \\
 \midrule
\multirow{3}{*}{macos-latest} & \multirow{3}{*}{macOS 14} & 18.20.8 & Google Chrome 136.0.7103.49 \\
 &  & 20.19.1 & Safari 18.4 (19621.1.15.111.1) \\
 &  & 22.15.0 &  \\
 \midrule
\multirow{3}{*}{windows-latest} & \multirow{3}{*}{Windows Server 2022} & 18.20.8 & Google Chrome 136.0.7103.93 \\
 &  & 20.19.1 & Microsoft Edge 136.0.3240.64 \\
 &  & 22.15.0 & Mozilla Firefox 138.0.1 \\ \bottomrule
\end{tabular}%
}
\end{table}

We manually customize the GitHub Actions workflow for each project to ensure compatibility and successful execution.
These adjustments help ensure that the test environment accurately reflects the project's expected setup and minimizes configuration-related failures that are unrelated to the environmental factors we vary.
The process includes setting the correct branch name when the default is not \texttt{main}, as some projects use alternative names such as \texttt{master} or custom branches for active development.
Additionally, we adapted the installation steps for each project to meet its specific requirements.
For instance, some projects require running custom scripts, using different package managers (e.g., \texttt{pnpm} vs. \texttt{npm}), or installing additional dependencies, such as system packages or global tools, before tests can run. 

After running each project across all environmental configurations, we saved the results of each rerun under the name of the corresponding forked project.
For some failing runs due to dependency errors or blocked GitHub Actions runs, we manually customize the workflow to fix dependency errors or failed tests.
For example, Bootstrap
requires JDK (Java) 17, but the default JDK in GitHub Actions on Windows Server 2022 is Java 8.
Therefore, we manually modify the workflow to install JDK 17.
After customizing the GitHub Actions workflow for each project, we commit the changes to our forked repository.
Pushing these changes triggers the GitHub Actions pipeline, which automatically starts running the test jobs.

\subsection{Manual Analysis}

We manually reviewed the results obtained from all \NumProjectsFinalDataset{} projects.
For every observed test failure in GitHub Actions, two of the co-authors independently examined the error message and traced the relevant code to determine the root cause.
If a failure was attributable to the OS, Node.js version, browser, or an interaction between them, we categorized it as \EnvFlakiness{} and assigned it to the corresponding category.
In cases where the two authors disagreed, a third author was involved to resolve the conflicts.
All disagreements were resolved through discussion until full consensus was reached.
In total, the two authors were able to reach consensus for \NumNoConflicts{} projects, and the third author needed to be involved for only \NumConflicts{} projects. 
In total, we observed test failures from \NumProjectsWithFailure{} projects in which \NumEnvProjects{} are \EnvFlakyProjects{} and \NumFlaky{} are flaky, caused by other factors, with four projects overlapping between the two categories.
\section{Results}
\label{sec:results}
\UseRawInputEncoding
We present the results in relation to the first two RQs below. 

\subsection{Prevalence of Environmental Flakiness in JavaScript (RQ1)}
By analyzing \NumProjectsFinalDataset{} projects, we categorise those projects into four groups based on the flaky tests we detected in those projects across different environmental configurations (i.e., OS type and Node.js version):

\begin{itemize}[leftmargin=*]
 \item \textbf{\NEDP{}}: projects that exhibit no test failures when running across different environmental configurations.

\item \textbf{\EDP{}}: projects with test failures caused by changes in the environment.
These are further divided into:
\begin{itemize}
\item \textbf{\EFP{}}: failures occur due to environment configurations at the project level, e.g., the project's package dependencies, before tests are even run.
\item \textbf{\EFT{}}: only specific tests fail under certain environment configurations, while the rest of the project remains stable.
\end{itemize}

\item \textbf{\FP{}}: projects that contain non-environment-related flaky tests beyond the factors that we control for, e.g., failures caused by network instability or randomness. \\
\end{itemize}

In total, from all \NumProjectsFinalDataset{} projects, we found \NumEnvProjects{} \EnvDepProjects{}, including \NumEnvFlaProjects{} \EnvFlakyProjects{} and \NumEnvFlaTests{} projects with \EnvFlakyTests{}.
We detect \NumEnvFlakyTests{} environment-dependent tests within those \NumEnvFlaTests{} projects.
On the other hand, \NumPassed{} projects are non-environment-dependent. \add{Most environmental flaky failures manifested on the first rerun after a change in the environment (77\% of projects). 
We required 10 reruns in the worst-case scenario.}

Table~\ref{tab:RQ1-results} indicates the number of projects in different categories, as well as the number of causes in each.
Note that a single project may belong to more than one group if it exhibits multiple types of behavior.
For example, a project may have tests that fail only in specific environments (categorized as \EnvFlakyTests{}) and also include tests that fail intermittently, regardless of the environment (categorized as \FlakyProjects{}).
As such, the summation of projects across all categories exceeds the total number of projects. 

Environmental flaky projects include different causes of flakiness: \NumEnvFlaProjectsOS{} are OS dependent, \NumEnvFlaProjectsNode{} are Node.js version dependent, and \NumEnvFlaProjectsBrowser{} are browser dependent.
In addition, the causes of test-level flakiness include \NumEnvFlaTestsOS{} OS dependent, \NumEnvFlaTestsNode{} Node.js-dependent, and \NumEnvFlaTestsBrowser{} browser-dependent flaky tests.

\begin{table}[!ht]
\centering
\caption{Projects with/without environmental dependencies}
\label{tab:RQ1-results}
\begin{threeparttable}
\begin{tabular}{p{2.1cm}rrrrr}
\toprule
            & \multicolumn{1}{l}{\#projects} & \multicolumn{1}{r}{\#NoD\tnote{*}} & \multicolumn{1}{r}{\#OSD\tnote{**}} & \multicolumn{1}{r}{\#BrD\tnote{***}} & \multicolumn{1}{>{\raggedleft}p{1cm}}{\#NoD \& OSD}\\\midrule
\NEDP{} & \NumPassed{}          & -     & -    & - & -  \\
\EFP{}  & \NumEnvFlaProjects{}  & 3     & \NumEnvFlaProjectsOS{}   & 8  & 8 \\
\EFT{}  & \NumEnvFlaTests{}     & 2     & 7   & 9  & 8\\
\FP{}   & 5           & -     & -    & -  & - \\\midrule
Total   & 122         & 5      & \NumEnvFlaOS{}     & \NumEnvFlaBrowser{}   & \NumEnvFlaOSNode{}   \\
\bottomrule
\end{tabular}
\begin{tablenotes}
  \item[*] Node dependent
  \item[**] OS dependent
  \item[***] Browser dependent
\end{tablenotes}
\end{threeparttable}
\end{table}
\vspace{-1.5mm}

In addition, \NumFlaky{} projects are flaky due to non-environment-related causes beyond those that we control for.
For example, Listing~\ref{lst:flakyEmxample} shows an asynchronous setup function 
that intermittently fails with the \texttt{error: Timeout - Async function did not complete within 10000ms (set by jasmine.DEFAULT\_TIMEOUT\_INTERVAL)}.
This failure occurs when the \texttt{await} operation, such as clearing the database, takes longer than Jasmine's default 10-second timeout, commonly due to slow database responses.


\begin{lstlisting}[caption={Failure caused by async wait},label={lst:flakyEmxample}]
beforeEach(async () => {
    await new MongoStorageAdapter({ uri: databaseURI }).deleteAllClasses();
    Config.get(Parse.applicationId).schemaCache.clear();});
\end{lstlisting}

\subsection{Root Causes of Environmental Flakiness (RQ2)}
By analyzing the failures observed across the \NumEnvProjects{} \EnvDepProjects{}, we investigate the root cause of the test failures and possible causes of test flakiness.
We categorized the detected flaky tests into four groups: (1)~OS dependency, (2)~Node.js version dependency, (3)~browser dependency, and (4)~combination of the above factors.


We next discuss the results concerning the four categories.
\subsubsection{OS dependency} 
Tests in this category fail when run on a specific OS due to differences in file system behavior (e.g., case sensitivity in file paths), line endings, shell command compatibility, or platform-specific dependencies.
%
In total, we identify \NumEnvFlaOS{} \EnvDepProjects{} with failures caused by running on a different OS.
Of these projects, \NumEnvFlaProjectsOS{} exhibit flakiness at the project level, while \NumEnvFlaTestsOS{} are \EnvFlakyTests{}.
Regarding failures on which OS, most failures occur on Windows (22 out of 26 projects), followed by macOS (8) and Ubuntu (2). Some projects exhibit failures on more than one OS.
\begin{lstlisting}[language=json,caption={Package.json using Unix-style cd commands},label={lst:osExp1}]
"test": "npm run --silent test:config && npm run --silent test:config:base",
"test:config": "cd packages/eslint-config-airbnb; npm test",
"test:config:base": "cd packages/eslint-config-airbnb-base; npm test",
\end{lstlisting}

An example of \textit{airbnb/javascript} 
with OS dependency is shown in Listing~\ref{lst:osExp1}.
The listing shows part of the \texttt{package.json} file from the Airbnb JavaScript Style Guide project
that fails on Windows. 
The build fails because the \texttt{test:config} and \texttt{test:config:base} scripts (lines 2 and 3) use Unix-style \texttt{cd} commands with semicolons (\texttt{;}) to chain commands.
On Unix-based systems, the semicolon acts as a command separator, but Windows does not recognize it in the same way.
As a result, the shell misinterprets the command and fails to execute the second part of the command.
This difference in interpretation leads to errors such as \textit{``The system cannot find the path specified''} and causes the test script to fail.
A cross-platform fix would involve replacing the semicolons with \texttt{\&\&}, which works consistently on both Unix-based systems and Windows.
\begin{lstlisting}[caption={OS-dependent test using file paths with backslashes},label={lst:osExp3}]
it('should output the correct snippet ids', () => {
  expect(snippetData.map(({ id }) => id).sort()).toEqual(
    [
      'articles/s/web-development-tips',
      'js/s/array-grouping',
      'js/s/array-initialize',
      'js/s/array-map-foreach',
      'js/s/array-compare',
      'css/s/content-centering',
      'css/s/css-reset',
    ].sort());});
\end{lstlisting}

Another example of path-specific issues is shown in Listing~\ref{lst:osExp3} (30-seconds-of-code).
The test fails on Windows because the OS returns file paths using backslashes (\texttt{\textbackslash}) instead of forward slashes (\texttt{/}).
This mismatch causes the test to fail when comparing the actual output to the expected snippet IDs.
To avoid such issues, the test code should normalize file paths using utilities such as \texttt{path.posix} or \texttt{path.normalize()} to ensure consistent comparisons across platforms.
\begin{lstlisting}[language=json,caption={Package.json using Unix-style cd commands},label={lst:osExp2}]
"scripts": {
    "test": "TZ=Pacific/Auckland npm run test-tz && TZ=Europe/London npm run test-tz && TZ=America/Whitehorse npm run test-tz && npm run test-tz && jest",
\end{lstlisting}

A similar issue occurs in the \textit{dayjs} project (Listing~\ref{lst:osExp2}), where the script fails on Windows due to platform-specific handling of environment variables.
The script sets the \texttt{TZ} (timezone) environment factor inline before calling \texttt{npm run test-tz} multiple times.
This syntax (\texttt{TZ=... command}) is available on Unix-based systems, where one can set environment variables inline.
However, this syntax is not natively supported on Windows, resulting in the test failing.
To fix this, tools like \texttt{cross-env}~\cite{npmCrossEnv} 
can be used to set environment factors in a way that works across different OS.

\subsubsection{Node.js Version Dependency} 
We identified \NumEnvFlaNode{} test failures that occur across different Node.js versions (i.e., a test passes in one Node.js version but fails in another).
Those failures are typically caused by breaking changes in the API, deprecated APIs, stricter error handling, or differences in built-in module implementations.
For example, a test may rely on undocumented behavior that changes between Node.js versions. 
In total, we identified four \EnvDepProjects{} with failures resulting from differences in Node.js versions.
Two of these projects are \EnvFlakyProjects{}, and two projects include \EnvFlakyTests{}, where only specific tests fail depending on the Node.js version.

Listing~\ref{lst:nodeExp1} shows a test from the \texttt{pug} repository that expects the error message to follow a specific structure.
The test passes when run with Node.js version 15 but fails when run with versions 18, 20, and 22 due to changes in the way error messages are generated~\cite{dawson2017node}. When running the test on version 18, the test shows an error message ``Cannot read property 'length' of undefined'', as it is expecting a \texttt{null} value.

\begin{lstlisting}[caption={Node-dependent test},label={lst:nodeExp1}]
it('includes detail of where the error was thrown including the filename', function() {
    var err = getFileError(
      __dirname + '/fixtures/runtime.with.mixin.error.pug',{});
    expect(err.message).toMatch(/mixin.error.pug:2/);
    expect(err.message).toMatch(/Cannot read property 'length' of null/);});
\end{lstlisting}
\Space{
\begin{lstlisting}[caption={Error message for a node-dependent test witherror message format check},label={lst:nodeExp11}]
    Expected pattern: /Cannot read property 'length' of undefined/
    Received string:  "/home/runner/work/pug/pug/packages/pug/test/fixtures/layout.with.runtime.error.pug:3
        1| html
        2|   body
      > 3|     = foo.length
        4|     block content
        5|.
    Cannot read properties of undefined (reading 'length')"
\end{lstlisting}}
    
Listing~\ref{lst:nodeExp2} presents another flaky test that passes on Node.js version 18 but fails on versions 20 and 22. In Node.js version 18, the ES module system is more forgiving, especially in test environments that transpile JSX before running the code.
However, Node.js versions 20 and 22 have stricter parsing rules and try to parse JSX as regular JavaScript, which causes syntax errors~\cite{nodejsV20ESMLoader}. 
To resolve this issue, the JSX code must be transpiled appropriately (e.g., using Babel.js transpiler) before running the tests, or the test runner configuration must be updated to handle JSX correctly in newer Node.js versions.
\begin{lstlisting}[caption={Node-dependent test using JSX},label={lst:nodeExp2}]
it('should return a valid Provider Component', () => {
    const { Provider } = createContext();
    const contextValue = { value: 'test' };
    const children = [<div>child1</div>, <div>child2</div>];
    const providerComponent = <Provider {...contextValue}>{children}</Provider>;    //expect(providerComponent).to.have.property('tag', 'Provider');
    expect(providerComponent.props.value).to.equal(contextValue.value);
    expect(providerComponent.props.children).to.equal(children);});
\end{lstlisting}

Of the projects we analyzed, we found that three of those projects declare a specific Node.js version or version range in their \texttt{package.json} file (via the \texttt{engines} field).
They were different from the versions that we considered for the running environment and caused failures.
We did not consider these failures as due to Node.js dependency issues.
This choice ensures that we do not report expected incompatibility errors as instances of environmental flakiness.
    
\subsubsection{Browser dependency}
\label{sec:browser}

Some failures typically occur in browser-based or end-to-end tests, where inconsistencies between browser engines (e.g., Chromium vs.\ WebKit) or browser versions lead to flakiness.
Errors can also be specific to certain combinations of browsers and OS (i.e., tests fail on specific browser setups within a specific OS). 
In total, we found 15 projects with browser-related flakiness, where \NumEnvFlaProjectsBrowser{} of them are flaky as projects, while the remaining \NumEnvFlaTestsBrowser{} have \EFT{}.
The majority of browser-related flakiness--observed in eight projects--occurs on macOS, affecting tests executed in Chrome, Firefox, or Safari.
These failures are typically caused by differences in browser engine behaviour, missing binaries, or variations in how browser APIs behave across platforms. 

Listing~\ref{lst:broExp1} shows an example of a configuration file from the \texttt{axios} project where part of the test suite is configured to run using Firefox.
On macOS, these tests fail with the error \textit{``No binary for FirefoxHeadless browser on your platform''}, as the required Firefox binary is not available in the test environment.
The error can be fixed by installing Firefox when running tests on macOS.
\begin{lstlisting}[language=json,caption={package.json using Firefox browser},label={lst:broExp1}]
"test:karma": "node bin/ssl_hotfix.js cross-env LISTEN_ADDR=:: karma start karma.conf.cjs --single-run",
"test:karma:firefox": "node bin/ssl_hotfix.js cross-env LISTEN_ADDR=:: Browsers=Firefox karma start karma.conf.cjs --single-run",
"test:karma:server": "node bin/ssl_hotfix.js cross-env karma start karma.conf.cjs",
"test:build:version": "node ./bin/check-build-version.js",
\end{lstlisting}

We can see a similar browser-related flakiness appears in the project \texttt{jQuery}, where a test fails exclusively on macOS (see Listing~\ref{lst:broExp2}).
The test, which checks the native abort behaviour of \texttt{jQuery.ajax()}, fails with the message ``unexpected success''.
This message indicates that the test passed when a failure was expected, suggesting a platform-specific difference in how \texttt{XMLHttpRequest} abort events are handled in Chrome's headless mode on macOS.
Such discrepancies can lead to unreliable outcomes in browser-based test suites across different environments.

\begin{lstlisting}[caption={Native abort behaviour},label={lst:broExp2}]
ajaxTest( "jQuery.ajax() - native abort", 2, function( assert ) {
  return {
    url: url( "mock.php?action=wait&wait=1" ),
    xhr: function() {
      var xhr = new window.XMLHttpRequest();
      setTimeout( function() {
        xhr.abort();
      }, 100 );
      return xhr;
    },
    error: function( xhr, msg ) {
      assert.strictEqual( msg, "error", "Native abort triggers error callback" );
    },
    complete: function() {
      assert.ok( true, "complete" );}};} );
\end{lstlisting}

    
\subsubsection{Combination of Node.js and operating system configurations} 
Flakiness in this category can be attributed to the interaction of multiple environmental factors.
For instance, a test may pass under most conditions but fail only on a specific OS with a particular version of Node.js or a browser.
Such issues are more complex to diagnose as they involve multiple combinations of environmental factors, making them difficult to reproduce.
Because of the number of cases and the importance of browsers in JavaScript, we consider the combination of browser and OS as part of the issue with browser, as discussed previously in Section~\ref{sec:browser}.

In this category, we identify a total of \NumEnvFlaOSNode{} projects with flakiness caused by the combination of OS and Node.js version.
Of these projects, \NumEnvFlaProjectsOSNode{} are \EFP{}, while \NumEnvFlaTestsOSNode{} are \EFT{}.
In addition, 13 of these projects fail on Windows, eight on macOS, and five failed from combinations involving Ubuntu and Node.js version.

An example is shown in the \textit{json-server} project (Listing~\ref{lst:comExp1}), where the tests fail on Windows when run with Node.js versions 18 and 20.
This failure transpired because, on Windows, the default command prompt (\texttt{cmd.exe}) does not perform glob pattern expansion.
As a result, commands like \texttt{node --test src/*.test.ts} pass the pattern \texttt{*.test.ts} as a literal string to Node.js, causing errors as the test runner cannot resolve the intended files.
Since version 22, an improved glob handling was introduced in the built-in test runner for Windows, which can mitigate this issue~\cite{nodejsIssue50658Comment}.

\begin{lstlisting}[language=json,caption={package.json file using command that not work on Windows with Node.js versions 18 and 20},label={lst:comExp1}]
"scripts": {
    "dev": "tsx watch src/bin.ts fixtures/db.json",
    "build": "rm -rf lib && tsc",
    "test": "node --import tsx/esm --test src/*.test.ts",
    "lint": "eslint src",
    "prepare": "husky",
    "prepublishOnly": "npm run build"},
\end{lstlisting}


\section{Mitigation Strategy (RQ3)}

\add{To address flaky tests due to environmental factors, we propose a lightweight mitigation approach that allows running specific tests or test files under certain environmental configurations, enabling CI builds to succeed without rerunning entire test suites due to flaky failures, thereby reducing developer inspection. We designed this approach to be intentionally lightweight so that it does not require rerunning the tests (which is expensive) and also requires minimal configuration.}

A previous study noted that when OS-specific conditions cause flakiness, developers often attempt to skip such OS-related flaky tests by embedding conditional logic to detect the current OS type and version within the test code, thereby allowing the tests to run only on specific OS~\cite{Hashemi2022flakyJS}.
\add{A common industrial practice to deal with flaky tests is to \textit{quarantine} them, i.e., allowing the CI to bypass the test while still reporting it as  flaky~\cite{amjed2022review,habchi2021qualitative}. Several test monitoring tools also support quarantining flaky tests by identifying them from previous runs~\cite{datadog2024flaky,atlas2025flaky,trunk2025flaky}.}
JUnit (for Java tests) provides a similar conditional test execution~\cite{junitConditionalExecution}
, which utilizes annotations to specify environmental conditions under which tests should (or should not) run.
\textit{@EnabledOnOs} and \textit{@DisabledOnOs} are JUnit annotations that, respectively, enable and disable tests on specific OS.
A GitHub code search shows more than 7,000 uses of \textit{@EnabledOnOs}~\cite{githubEnabledOnOsSearch} 
and more than 6,000 uses of \textit{@DisabledOnOs}~\cite{githubDisabledOnOsSearch}. 
A similar approach has also been adopted in a previous tool, \textit{saflate}, which sanitizes flaky tests due to network dependencies (i.e., tests that fail due to network availability exceptions in JUnit) using custom annotations added to test files or tests~\cite{dietrich2022flaky}. 

Our approach employs a similar strategy for JavaScript tests, aiming to develop an expandable, cross-framework solution that works across popular JavaScript testing frameworks. It introduces comparable functionality through static code transformation, enabling developers to skip tests based on simple annotations in their test files. \add{It also generates detailed reports explaining why tests were skipped and under what conditions. These reports describe the affected tests and environments, providing developers with a clear starting point for diagnosing and ultimately fixing the underlying issue.}

\subsection{Approach}

The primary goal of our lightweight approach to address environment-specific flaky tests is to allow developers to manage them using annotations.
Developers annotate their tests with docblock~\cite{adobeJsDocblock} 
tags, specifying the conditions under which a test should be skipped or enabled.
This approach eliminates the need for ad-hoc conditional logic within test code, promoting cleaner and more maintainable test suites.
All intentionally skipped tests are then reported in a generated test report, which shows which tests are skipped and the identified possible reason for the flakiness. 

The approach is designed to support multiple JavaScript testing frameworks, without requiring modifications to existing test logic. The core components of the proposed approach are:

\begin{itemize}
    \item \textbf{Annotation Parser:} Extracts metadata from test file docblocks, identifying \textit{skip} or \textit{enable} conditions.
    \item \textbf{Environment Detector:} Detects, at runtime, the current operating system, Node.js version, and the active browser.
    \item \textbf{Condition Evaluator:} Compares the extracted annotations with the detected environment to decide whether a test should be skipped.
    \item \textbf{Test Skipping Module:} Marks tests as skipped and logs skip reasons for review, before test execution.
\end{itemize}

The high-level workflow of the tool is as follows: when a test file is loaded, it parses all docblock annotations, evaluates them against the runtime environment, and automatically skips tests that do not meet the specified conditions.
This design ensures consistent and automated enforcement of environment-specific constraints across different testing frameworks. 

We have developed a set of annotation tags that can be used to enable/disable tests on specific environment configurations: OS types, Node.js versions, and browser types.
Table~\ref{tab:ann-tags} lists all the annotation tags we defined, along with their descriptions.
Tag names and values are case-insensitive, and lists are comma-separated.
Each test will run as the default unless a \texttt{skip} condition matches. 



\begin{table}[!ht]\centering
\caption{Annotation tags supported by \tool{}}
\label{tab:ann-tags}
\resizebox{\columnwidth}{!}{
\begin{tabular}{@{}llp{5cm}@{}}
\toprule
\textbf{Type} & \textbf{Tag} & \textbf{Effect when condition holds} \\ 
\midrule

\multirow{2}{*}{OS} 
  & \texttt{@enableOnOs}  
  & \multirow{2}{=}{Run only (or skip) on the specified OS.} \\
  & \texttt{@skipOnOs} & \\ 
\midrule

\multirow{4}{*}{Node.js} 
  & \texttt{@enableOnNodeVersion} 
  & \multirow{2}{=}{Run only (or skip) on the specified Node.js version(s).} \\
  & \texttt{@skipOnNodeVersion} & \\ 
  \cmidrule(lr){2-3}
  & \texttt{@enableOnNodeRange} 
  & \multirow{2}{=}{Run only (or skip) when Node.js satisfies the given version range.} \\ 
  & \texttt{@skipOnNodeRange} & \\ 
\midrule

\multirow{2}{*}{Browser} 
  & \texttt{@enableOnBrowser}  
  & \multirow{2}{=}{Run only (or skip) when the active browser matches one of the specified values.} \\
  & \texttt{@skipOnBrowser} & \\ 

\bottomrule
\end{tabular}}
\end{table}
\vspace{-2.5mm}

\subsection{Implementation}
We implemented our lightweight sanitization tool, \tool{}, as a Babel plugin\footnote{\url{https://babeljs.io/docs/plugins}}, extending the test runner to intercept the execution pipeline before test execution.
It processes the test suite by scanning the docblock annotations associated with each test.
These annotations define constraints 

The steps taken to run or skip tests are detailed in Algorithm~\ref{alg:env-skip}.
At first, \tool{} extracts the current runtime environment (i.e., OS, Node.js version, and the browser).
Each test's annotations are then parsed and compared against this environment.
If the conditions indicate that a test should not run (e.g., \texttt{@skipOnOS win32} when the OS is Windows), the test is marked as such in a test log saved in the project directory.
Otherwise, the test is preserved for execution. This approach ensures that environment-specific tests are automatically skipped without manual intervention, improving test stability and reproducibility.
The tool operates without modifying the original test code, 
facilitating easy integration into CI workflows. 

The tool currently supports three testing frameworks: Jest, Mocha, and Vitest.
However, the tool is extensible and can be readily expanded to support other testing frameworks.

\begin{algorithm}[]
\caption{Test sanitization using annotations}\label{alg:env-skip}
\begin{algorithmic}[1]
\State \textbf{Input:} Test suite
\State \textbf{Output:} Modified ineligible test blocks marked as skipped
\State Load test suite
\State Detect current OS, Node.js version, and browser (if applicable)
\For{each testing block in test suite}\Comment{describe, it or test blocks}
    \If{\textit{it has supported annotation tags before it}}
            \Statex \hspace{\algorithmicindent} \textit{shouldSkip} $\gets$ !~(environment matches all enable-conditions) $\land$ (environment matches any skip-condition)
    \EndIf
    \If{\textit{shouldSkip}}
        \State Change the state of the test to skip
        \State Log skip reason with timestamp and matched condition
    \EndIf
\EndFor
\State Return test suite for execution
\end{algorithmic}
\end{algorithm}
\vspace{-2mm}

\subsection{Evaluation}
We evaluate \tool{} using a subset of projects from our dataset (discussed in Section~\ref{sec:dataset})  that (1)~use one of the supported test frameworks (Jest, Mocha, or Vitest) and (2)~contain at least one environmental flaky test.
We start with \NumEvaluationProjects{} candidate projects.
Among these projects, \NumMixFrameworks{} use multiple testing frameworks or runners, and in those cases, we found that the environmental flaky tests occur in unsupported frameworks.
For example, \texttt{swagger-ui} uses both Jest and Cypress tests.
Upon investigation, we found that the two detected environmental flaky tests occurred in the Cypress test suite, which \tool{} does not currently support.
We therefore exclude those \NumMixFrameworks{} projects from the evaluation and report results on the remaining projects.
The projects we included in our evaluation are three projects that use Jest, four that use Mocha, and two that use Vitest.

By declaring \tool{} under a project's \texttt{devDependencies}, the package auto-wires itself during installation and activates at test execution time without any manual setup.
It detects the host test runner (Jest, Mocha, or Vitest) and attaches the appropriate lightweight integration (e.g., \emph{babel-jest} transform, a minimal \emph{@babel/core} require-hook for Mocha, or a vite Babel plugin for Vitest). 

For each project, we add \tool{} as a dependency into the package and apply the appropriate annotation based on the experiments described in Section~\ref{sec:results}.
We then push these changes to the forked projects and run the project's CI workflow, which executes the test suite across nine environment configurations, each repeated ten times. 

In two projects, \textit{insomnia} and \textit{mongoose}, we were unable to evaluate this sanitization approach.
\textit{insomnia} features multiple subprojects and workspaces that share a common \texttt{Vitest} configuration.
This made the instrumentation more challenging, as the test configuration file is not localized to a single package but instead inherited across different workspaces.
Consequently, our setup script could not fully integrate with the heterogeneous project setup, and some tests executed before the sanitizer hooks could not be properly run.
Similarly, \textit{mongoose} uses\texttt{Mocha}, and most of its tests require a live \textit{MongoDB} instance for execution. 
\textit{mongoose} already includes its own mechanism to skip specific database-dependent tests when a required environment variable (\texttt{MONGOOSE\_SHARD\_TEST\_URI}) is not set (an interestingly similar sanitization approach to ours that targets database configurations).
However, once we added \tool{}, this built-in skipping mechanism did not take effect as expected, and the connection attempts were still triggered.
In practice, the tests that should have been skipped by \textit{mongoose}'s own logic were executed anyway, leading to failures.
Therefore, we excluded the two projects from our evaluation and reported the results for the other \NumEvaluationProjectsFinal{} projects.

Table~\ref{tab:evaluation} summarizes the evaluation results.
Shaded cells indicate the number of test failures due to environmental dependency, followed by the number of skipped tests after applying our sanitization tool.
The detailed evaluation results are available in our replication package~\cite{replicationPackage}.
As shown in Table~\ref{tab:evaluation}, the results show that \tool{} can successfully sanitize and skip all tests that have failed due to environmental (OS, Node.js, or browser) dependencies.
\add{In other words, the initially failing tests have all been successfully quarantined and reported as skipped, allowing the CI build to continue without the known test failure.}

\add{To evaluate the tool's accuracy, we checked for any false positives (FPs), i.e., tests incorrectly skipped, and false negatives (FNs), i.e., tests that should have been skipped but were not. Based on our manual verification, we did not observe any FPs or FNs, indicating that the tool has a high accuracy.
The tool is also shown to have a low performance overhead (see the runtime data in Table~\ref{tab:evaluation}; runtime values are averages across 90 runs).
For most projects, we observe no significant difference in average runtime before and after sanitization.
A notable exception is \textit{fabric.js}, where the average runtime drops significantly as skipping the tests allowed the CI to continue, reducing the overall runtime.} 

Listings~\ref{lst:expSkipPreact} 
demonstrate the use of annotation tags to skip \EnvFlakyTests{}, corresponding to the examples in Listings~\ref{lst:nodeExp2} and \ref{lst:osExp3} in Section~\ref{sec:results}, respectively.
The annotation tag ``\texttt{@skipOnNodeVersion 20,22}'' restricts execution of the test to Node.js version 18, and automatically skips the test on versions 20 and 22, where it is known to fail.
Similarly, the annotation tag ``\texttt{@skipOnOS win32}'' ensures that the test is skipped on Windows, preventing OS-specific failures (due to the use of Unix-like paths) in Windows. 

\begin{lstlisting}[caption={Annotation tags to skip tests on Node and Windows},label={lst:expSkipPreact}]
/**
  * @skipOnNodeVersion 20,22
  */
it('should return a valid Provider Component',

/**
  * @skipOnOS win32
  */
it('should output the correct snippet ids',
\end{lstlisting}


\begin{table*}[!ht]\centering
\caption{Results of sanitizing environmental flaky tests using \tool{}}
\label{tab:evaluation}
\begin{tabular}{@{}llrrrrrl|rrrr@{}}
\toprule
 &  & \multicolumn{1}{l}{} & \multicolumn{4}{c}{\textbf{Default tests  run}} &  & \multicolumn{4}{c}{\textbf{Sanitized Tests}} \\ \cmidrule(l){4-12} 
\multirow{-2}{*}{\textbf{Framework}} & \multirow{-2}{*}{\textbf{Project}} & \multicolumn{1}{l}{\multirow{-2}{*}{\textbf{\#tests}}} & \multicolumn{1}{l}{\cellcolor[HTML]{EFEFEF}\#failure} & \multicolumn{1}{l}{\#skipped} & \multicolumn{1}{l}{\#passed} & \begin{tabular}[c]{@{}r@{}}Avg \\ runtime (s)\end{tabular} & Cause & \multicolumn{1}{l}{\#failure} & \multicolumn{1}{l}{\#passed} & \multicolumn{1}{l}{\cellcolor[HTML]{D7E0E0}\#skipped} & \begin{tabular}[c]{@{}r@{}}Avg \\ runtime (s)\end{tabular} \\ \midrule
Jest & iptv & 8 & \cellcolor[HTML]{EFEFEF}8 & 0 & 7 & 31 & OS \& Node & 0 & 8 & \cellcolor[HTML]{D7E0E0}8 & 77 \\
 & fabric.js & 162 & \cellcolor[HTML]{EFEFEF}99 & 2 & 61 & 1020 & OS \& Node & 0 & 61 & \cellcolor[HTML]{D7E0E0}101 & 65 \\
 & pug & 650 & \cellcolor[HTML]{EFEFEF}2 & 0 & 648 & 59 & Node range & 0 & 648 & \cellcolor[HTML]{D7E0E0}2 & 85 \\
 \midrule
Mocha & rollup & 5213 & \cellcolor[HTML]{EFEFEF}1 & 0 & 5212 & 718 & OS \& Node & 0 & 5212 & \cellcolor[HTML]{D7E0E0}1 & 721 \\
 & sails & 1550 & \cellcolor[HTML]{EFEFEF}1 & 13 & 1536 & 237 & Node & 0 & 1536 & \cellcolor[HTML]{D7E0E0}14 & 238 \\
 & preact & 1217 & \cellcolor[HTML]{EFEFEF}1 & 13 & 1203 & 68 & Node & 0 & 1203 & \cellcolor[HTML]{D7E0E0}14 & 147 \\ \midrule
Vitest & 30-seconds-of-code & 148 & \cellcolor[HTML]{EFEFEF}17 & 0 & 131 & 40 & OS & 0 & 131 & \cellcolor[HTML]{D7E0E0}17 & 63 \\ \bottomrule
\end{tabular}%
\end{table*}
\vspace{-2mm}

\section{Discussion}
\label{sec:discussion}


Our systematic evaluation of environmental factors reveals that environmental flakiness is common in JavaScript projects and can manifest in various ways.
For example, the whole project may fail due to environment-specific settings (at the package level), or only a few tests break under specific conditions.
We have identified tests that can run under specific steps, such as with a particular OS, Node.js version, or browser (or a combination of these factors).
Based on our results, many OS-dependent issues were specific to Windows, with path-related issues being the most common type.
Browser-related problems mainly appeared when running the test in macOS. 

Environmental flakiness can reveal deeper issues, such as features that are not equally supported across platforms or code that depends on OS-specific behavior. Developers often need to run tests in multiple environments, and when failures occur, they should reproduce the problem in the same environment where it originally happened. Our approach encourages cross-environment testing, allowing developers to control which tests can be run across different environmental setups without the need to change the test logic. 

The root causes we identified point to practical lessons.
For OS-dependent issues, cross-platform practices such as using \texttt{cross-env} for environment variables or normalizing file paths make tests more reliable.
For Node.js-dependent issues, it helps to clearly specify supported versions in \texttt{package.json} and avoid fragile assertions that depend on exact error messages.
For browser-dependent failures, making sure the right browser binaries are installed in CI and writing tests that work across different engines reduces surprises.

\section{Threats to validity}
\label{sec:threats}

The study uses open-source JavaScript projects available on GitHub.
The observations regarding environmental flakiness are limited to the selected projects and may not be fully generalizable to all JavaScript projects.
To mitigate the threat of generalizability, we selected popular (ranked by stars) repositories maintained by large communities and spanning diverse application domains, including servers, front-end and back-end frameworks, and web applications.
The selected projects also use a variety of testing frameworks. 

Part of the project selection and the analysis of the cause of flakiness was performed manually, which may introduce construct validity threats due to potential oversight or bias.
The manual analysis consisted of two stages: (1)~installing and investigating the tests in each project to observe and classify the impact on environmental flakiness, and (2)~installing \tool{} and manually adding annotation tags to skip the \EnvDepTests{} under the corresponding environments.
To mitigate bias, two authors independently examined the source code and error messages to reduce false positives in classification.
In cases of disagreement, a third author independently reviewed the instance, and all three authors discussed the causes until consensus was reached.
\section{Related Work}
\label{sec:related}

A large number of studies investigated various aspects of flaky tests, including root causes and methods for identifying, detecting, predicting, or mitigating them~\cite {thorve2018empirical,eck2019understanding,amjed2022review}.
Previous studies have focused mainly on specific causes of test flakiness, such as test-order dependency~\cite{gambi2018practical,hashemi2025detecting, lam2019idflakies}, concurrency~\cite{rahman2024flakesync,rahman2024flakerake,dong2020concurrency}, unordered collections~\cite{shi2016nondex}, code instrumentation~\cite{rasheed2023effect}, or network flakiness~\cite{dietrich2022flaky}.

\subsection{Empirical Studies on Test Flakiness}
Luo et al.~\cite{luo2014empirical} reported the first empirical study on flaky tests in Java projects.
The study classified the root causes of flaky tests into ten main categories, noting that one of these causes is environment dependency.
Eck et al.~\cite{eck2019understanding} also cited \textit{platform dependency}, which describes flaky behavior that occurs only on specific platforms (OS or virtual environment). 
 
Hashemi et al.~\cite{Hashemi2022flakyJS} examined root causes and how developers attempt to address flaky tests in JavaScript.
One of the main causes reported is OS-dependency, in which a test passes on one OS but fails on another.
The study reported that developers usually choose to skip, ignore, or disable the problematic tests by adding conditions to the tests that check the underlying OS before executing them.
Similarly, Job et al.~\cite{job2025and} studied OS-specific tests in Python,\Space{examining how and why developers implement them,} and found that OS-specific tests are common (56\% of the analyzed projects), with Windows-specific tests being the most common type.

\subsection{Flakiness Caused by Environmental Factors}
Environmental factors, such as OS, external libraries, or compilation environments, can significantly influence test behavior, often leading to non-deterministic outcomes~\cite{micco2017google,terragni2020container}.
Such variability is a common source of flakiness, as tests that pass in one environment may fail in another due to subtle differences in system behavior or resource availability.
Developers frequently introduce OS-specific tests to account for these variations \cite{Hashemi2022flakyJS}, performing operations such as calling platform-dependent APIs, mocking OS-level objects, or controlling execution flow. 
Job et al.~\cite{job2025and} show that OS-specific tests are often employed to work around missing external resources, unsupported standard libraries, and flaky tests.


A study on SAP HANA's CI pipeline~\cite{berndt2024test} emphasizes the significant impact of the testing environment on test flakiness.
It has been reported that environmental factors, including the execution infrastructure and test distribution, play a significant role in producing flaky behavior.
While higher test execution times, possibly due to limited infrastructure speed, correlated with increased flakiness, other factors such as system load, memory, and CPU usage did not show a clear relationship. 

FlakyLoc~\cite{moran2019debugging} aims to identify the root causes of flakiness in web applications by executing known flaky tests across different environmental configurations (e.g., memory size, browsers, and screen resolution), which vary across executions.

\subsection{Flaky Tests in JavaScript}
An empirical study of flaky tests in JavaScript projects identified concurrency, async wait, and OS as the top three causes of test flakiness~\cite{Hashemi2022flakyJS}.
Regarding how flaky tests are handled, the majority of such tests (> 80\%) are fixed by developers to eliminate flaky behavior.
Costa et al.~\cite{costa2022test} investigated flaky tests across five languages, including JavaScript, and found that \emph{async wait}, \emph{concurrency}, and \emph{platform dependency} were the top causes of flakiness in JavaScript. 

Heshemi et al.~\cite{hashemi2025detecting} investigated order-dependent tests in JavaScript by systematically reordering and rerunning tests and test suites.
The study found that order-dependent tests are caused by either shared file or shared mocking states between tests. 
Yost~\cite{yost2023finding} employed a combination of stress testing and test suite reordering to detect flaky tests in JavaScript applications.
The proposed method involved subjecting the system to stress and altering the execution order of test cases to expose issues that may remain undetected under standard conditions. 
FlakyFix~\cite{fatima2024flakyFix} utilized large language models to classify the types of fixes applied to flaky tests.
The proposed approach examines test code to uncover patterns that contribute to flakiness, enabling the prediction of fix categories. 

There are only a few dedicated flaky test tools 
for JavaScript~\cite{parry2021survey,amjed2022review}. \Space{Of those tools,
King et al.~\cite{king2018towards} proposed a machine learning-based approach that conceptualizes flakiness as a condition with recognizable symptoms, aiding in uncovering its root causes. Their tool uses Bayesian networks to classify and predict flaky tests in JavaScript.} Of those tools,
NodeRacer~\cite{endo2020noderacer} focuses on detecting flakiness due to concurrent, event races in Node.js applications.
JS-TOD \cite{hashemi2025js} focuses on detecting order-dependent flaky tests in JavaScript by systematically reordering and rerunning tests. 

\section{Conclusion}
\label{sec:conclusion}

\Space{Test flakiness poses a significant challenge to the test and product quality.
JavaScript is no exception to the possible impact of flakiness on test reliability and correctness.
One of the less investigated root causes of flakiness is the environmental factors under which the tests (or the project as a whole) run.} 
In the study, we present an in-depth systematic evaluation of environmental factors in JavaScript projects. \Space{We collected data from \NumProjectsFinalDataset{} project from GitHub, and analyzed possible flaky behavior caused by variation in the test environment, considering three environmental factors: OS type, Node.js version, and browser type.} 
We analyzed possible flaky behavior caused by variation in the test environment, considering three environmental factors: OS type, Node.js version, and browser type.
Our evaluation shows that OS-dependency is the most common type of environmental flakiness and happens across testing frameworks (\NumEnvFlaOS{}).
This includes issues related to path formatting (between Windows and Unix-like systems) and the use of OS-specific options.
There are also other issues related to Node.js versions (i.e., the use of unsupported features in older versions) or the use of a specific browser in a specific OS.

To address these issues, we developed \tool{}, a lightweight annotation-based sanitization tool that supports Jest, Mocha, and Vitest testing frameworks.
The tool allows developers to define tests that run only on a specific environment, enabling tests to skip if the environment differs from the defined configuration.
This approach allows CI builds to continue and succeed without rerunning entire test suites, while also reporting and documenting the intentionally skipped tests for future fixes.


\bibliographystyle{IEEEtran}
\bibliography{bibliography}

\end{document}